\documentclass[12pt]{article}
\input epsf.tex
\def\gs{g_{\rm{s}}}
\def\ls{\ell_{\rm{s}}}
\def\gYM{g_{\rm{SYM}}}
\def\Ue{U_{\rm e}}
\def\lra{\leftrightarrow}
\def\ba{\begin{array}}
\def\ea{\end{array}}
\def\be{\begin{equation}}
\def\ee{\end{equation}}
\def\bs{\vspace{5pt}}
\def\half{\mbox{\scriptsize{${{1}\over{2}}$}}}
\thicklines
\begin{document}
\begin{flushright}
\begin{minipage}{0.25\textwidth}
NSF-ITP-00-18\\
hep-th/0003251
\end{minipage}
\end{flushright}
\begin{center}

\bigskip\bigskip\bigskip

{\bf\large{Excision of `repulson' singularities: \\
a spacetime result \\ and its gauge theory analogue\\}}

\bigskip

{\em Amanda W. Peet\footnote{{\tt peet@itp.ucsb.edu}}\\ Institute for
Theoretical Physics,\\ University of California,\\ Santa Barbara, CA
93106-4030, U.S.A.}

\bigskip

Based on [1] with C.V.~Johnson and J.~Polchinski

\bigskip\bigskip\bigskip

\begin{minipage}[c]{0.8\textwidth}

We discuss spacetime singularity resolution in the context of the
gravity / gauge correspondence, for brane systems which give rise to
gauge theories with eight supercharges and no hypermultiplets.  The
discussion is aimed at non-experts.  Writeup of talk on [1] for
proceedings of PASCOS-99 (10 - 16 Dec 1999, Granlibakken, Lake Tahoe);
also relevant to Way Beyond the Standard Models Conference (31 Jan - 5
Feb 2000, Aspen, Colorado) and CIAR Gravity and Cosmology Programme
Meeting (19 - 22 Feb 2000, Banff, Canada).

\end{minipage}
\end{center}

\bigskip\bigskip\bigskip
\section{Introduction}

Spacetimes with singularities occur in many classical theories of
gravity.  This includes classical superstring theory
i.e. supergravity, where there are examples of spacelike, null, and
timelike singularities, some of which are naked.  An important
question is how, or indeed whether, quantum string theory resolves
these singularities.

As pointed out in Ref.~\cite{gthrcm}, some singularities are unphysical
and cannot be patched up by string theory.  A canonical example is the
negative-mass Schwarzschild geometry which has a naked singularity.
If string theory were to patch that up by smoothing it out to a small
region of strong but finite curvature, then by mass negativity this
would signal an instability of the theory as the vacuum would be
unstable.

String theory resolves some spacetime singularities in a beautiful
way.  The basic idea is that once spacetime curvatures or other
invariant measures of quantum corrections such as the dilaton are too
large for supergravity to be applicable, other degrees of freedom take
over.  Examples of this are the gravity / gauge theory correspondences
with sixteen supercharges of Ref.~\cite{IMSY}.  Let us now turn to an
introduction to these correspondences; a comprehensive review of the
subject including the original $AdS$/CFT correspondences of
Ref.~\cite{malda1} may be found in Ref.~\cite{agmoo}.

\begin{figure}
\epsfysize=1.0truein
\hskip0.3\textwidth\epsfbox{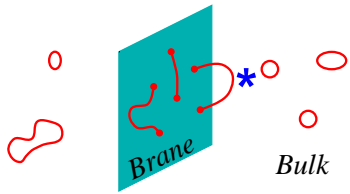}
\caption{\small A D$p$-brane living in the $d\!=\!10$ bulk, with 
open and closed strings. The $\star$ represents open-closed
interactions.}
\label{fig:decoupling}
\end{figure}

The starting point for correspondences with sixteen supercharges is a
system of $N$ D$p$-branes of Type II string theory.  These D$p$-branes
are hypersurfaces where open strings must end.  The open strings
distinguish which brane they end on via their Chan-Paton factors,
which they carry on their endpoints at zero energy cost.  This gives
rise to a $U(N)$ gauge theory living on the worldvolume of the
D$p$-branes.  In addition, the D$p$-branes carry mass and
Ramond-Ramond charge in the ten-dimensional bulk due to interactions
between open and closed strings.  See Ref.~\cite{Joebook} and
Fig.\ref{fig:decoupling}.

To engineer gravity/gauge theory correspondences, one takes a special
low-energy limit of the D$p$-brane system.  In the limit, the
complicated physics on the branes reduces to $d\!=\!p\!+\!1$
supersymmetric $U(N)$ Yang-Mills theory (SYM) with sixteen
supercharges.  For $p\!>\!3$, the SYM theory is nonrenormalizable and
so there is a need for new ultraviolet degrees of freedom in order to
define the theory.  String theory provides them in a well understood
fashion for $p\!<\!6$.  We will not discuss the $p\ge{6}$ cases here.

In the limit, gauge theory energies $E$ are taken to be well below the
string scale, $E\ls\rightarrow 0$.
In order to keep the brane physics nontrivial, the gauge coupling on
the branes, which is a derived quantity,
$\gYM^{2\,(p)} = (2\pi)^{p-2} \gs \ls^{p-3}\,,$
is held fixed.  Here $\gs$ is the string coupling constant.  BPS
W-boson-like states in the gauge theory, which in the bulk picture are
open strings stretched perpendicularly between D$p$-branes, have mass
$U\equiv r\ls^{-2}$
which is also held fixed.  In this relation, $r$ is the separation
between the D$p$-branes.  

In string theory the bulk Newton constant is also a derived quantity,
e.g. in $d\!=\!10$ we have
$G_{10} = 8\pi^6 \gs^2\ls^8 \,.$
The D$p$-brane supergravity geometry may be written in terms of a
harmonic function $H_p = 1 + c_p \gs N (\ls/r)^{7-p}$, where $c_p$ is
a constant.  Use of the above scaling relations results in the loss of
the $1$ from $H_p$, i.e. the asymptotically flat part of the geometry.
As a result, the bulk description of the system of $N$ D$p$-branes in
the above low-energy limit is in terms of string theory on the
near-horizon geometry.

This limit is called the decoupling limit because at such low energies
the coupling between open and closed strings is turned off.  Hence,
the gauge theory on the branes and the bulk string theory on the
near-horizon geometry may each be considered as a unitary theory on
their own.  This property makes a duality between the two theories
feasible.  Such a duality is often termed `holographic' because it
relates a $d\!=\!10$ bulk string theory to a $d\!=\!p\!+\!1$ gauge
field theory.  Note also that in this limit, there is no gravity on
the brane, unlike the situation that occurs in the Randall-Sundrum
scenario of Ref.s~\cite{lisaraman}.  

For general $p$, the spacetime curvature of the D$p$-brane
near-horizon geometry is not constant but varies with radial variable
$U$.  The local value of the string coupling, $\gs e^\Phi$, also
varies, but differently.  For example, the cases $p\!<\!3$ have a
dilaton singularity at $U\!\rightarrow\!{0}$ and a curvature
singularity at $U\!\rightarrow\!\infty$, while for $p\!>\!3$ cases it
is the other way around.  We then need to know what degrees of freedom
take over where the bulk $d\!=\!10$ supergravity description goes bad.
In regions where the dilaton is strong, we use S-duality which takes
us to a description with a weak dilaton.  In regions where the
curvatures become strong, it turns out that the right description is
the SYM theory.  For details, see Ref.~\cite{IMSY} and further
development reviewed in Ref.~\cite{emil}.  Before moving on, we will
just note that by the nature of duality, only one description may be
weakly coupled in any given region of parameter- and $U$- space.

If we wish to do gauge theory at finite temperature $T$, we use the
nonextremal D-brane geometry with Hawking temperature $T$, taken in
the decoupling limit.  This gives rise to a black hole type
generalization of the near-horizon geometry.  In principle, this is a
useful picture in the context of the black hole information problem,
because we have a duality of the black hole type system to a quantum
field theory which is a manifestly unitary theory.  The difficulty in
solving the information problem in practice is the strong-weak nature
of the duality: when the bulk picture is weakly coupled (calculable),
the brane picture is strongly coupled and vice-versa.

There have been many further generalizations of the prototype
correspondences; see Ref.s~\cite{agmoo,giveonkutasov}.  We note here,
however, that some attempts to construct supergravity duals to certain
gauge theories turned out to be unphysical, partly because analyzing
the supergravity theory in situations of interest is generally very
difficult.  See e.g. Ref.~\cite{goodbadnaked} for a discussion of some
of the problems encountered.

\section{The pure ${\mathbf{\cal{N}}}=2$ correspondence}

We are interested in a system which gives pure gauge theory with eight
supercharges living on the branes.  Examples with eight supercharges
and lots of hypermultiplet matter included the original
$AdS_3{\times}S^3{\times}T^4$ and $AdS_3{\times}S^3{\times}K3$
correspondences and Ref.s~\cite{igornikitaigorarkadyum}.  However, we
want no hypermultiplets, and this will have drastic consequences for
the properties of our spacetime.

Our setup of Ref.~\cite{JPP} includes many dual realizations in terms
of branes.  The one on which we will concentrate is obtained by
wrapping D$(p\!+\!4)$-branes on $K3$.  Other options including
D$(p\!+\!1)$-branes strung between two parallel NS5-branes, and
D$(p\!+\!2)$ wrapped on collapsing S$^2$ cycles in $K3$.  Related work
appeared in yet another dual realization, the heterotic one, in
Ref.~\cite{morten}.

\subsection{The setup}

Here we will focus on the $p\!=\!2$ case for simplicity.  Our starting
point is therefore a system of $N$ D6-branes wrapped on a $K3$
surface.

For the gauge theory side, we build the $d\!=\!2\!+\!1$ 't Hooft
coupling out of the $d\!=\!6\!+\!1$ coupling and the $K3$ volume
$(2\pi R)^4$:
\be
\lambda_2 \equiv g^2_{\rm SYM, 2} N 
= {{g^2_{\rm SYM, 6}N}\over{(2\pi{R})^4}} = {{\gs N \ls^3}\over{R^4}} \,.
\ee
The gauge theory will indeed be $2\!+\!1$ dimensional as long as
excitations are low-energy by comparison to the characteristic scale
of excitations in the compact $K3$.  We therefore keep $1/R$ finite in
the decoupling limit.

For the bulk side of things, in constructing the supergravity solution
we usually start by finding the conserved charges.  Generally, a
supergravity solution is then uniquely determined via a no-hair
argument.  It is important to realize, however, that this argument
fails when the geometry has a naked singularity, and so we will have
to proceed with caution if we encounter one.

We begin with $N$ D6-branes, so we have D6-brane charge $N$.  Once we
wrap a D6 on a $K3$, an additional charge arises due to the curvature
of $K3$, see Ref.s~\cite{suniloneBBG}.  As a result, we also have
D2-brane charge $(-N)$.  The ADM tension formula is protected by
supersymmetry and so we have
\be
\tau_{\rm ADM} = {{N}\over{(2\pi)^3\gs\ls^7}} \,R^4
\,{{-}}\, {{N}\over{(2\pi)^3\gs\ls^3}} \,.
\ee
Applying the no-hair theorem naively, we find that the metric the
string sees is, in the decoupling limit,
\be
\ba{ll}\label{repulsonmetric}
\bs\bs
{\displaystyle{{dS^2}\over{\ls^2}}} = &\! {\displaystyle{ 
{{R^{-2}dx_\parallel^2}\over{\sqrt{(\lambda_2/U)[1-(\lambda_2/U)]}}} 
  + \sqrt{{[1-(\lambda_2/U)]}\over{(\lambda_2/U)}} ds^2_{K3} }} \cr
{\ } &\!+ 
R^2\sqrt{(\lambda_2/U)[1-(\lambda_2/U)]}
\left\{dU^2+U^2\,d\Omega_{2}^2\right\} \,. 
\ea
\ee
We have used spherical coordinates $(U,\Omega_2)$ for the three
dimensions transverse to both the $K3$ and the $d\!=\!2\!+\!1$
worldvolume $x_\parallel$. 

Let us inspect our classical metric.  At
\be
U = \lambda_2 \,,
\ee
there is a pathology; some components of the metric become imaginary
for $U\!<\!\lambda_2$.  Computing curvature invariants, we find that
there is a singularity at $U\!=\!\lambda_2$, the locus of which is a
two-sphere.  The would-be horizon, located where $g^{UU}\rightarrow
0$, is at $U=0$, and so the singularity is naked.

In a related (heterotic) context, this singularity was studied in
Ref.s~\cite{rep1,rep2}, and in Ref.~\cite{rep2} it was dubbed the
`repulson' because massive particles are repelled by the singularity.
This behavior is reminiscent of the inside of extremal
Reissner-N{\"{o}}rdstrom black hole.

Given our previous caveat on use of no-hair theorems, and the fact
that some classical spacetime singularities are unphysical and cannot
be resolved by quantum stringy effects, we have cause for concern
about our repulson.  In fact, it turns out that the repulson
singularity is excised via a stringy mechanism, and we now turn to the
description of this excision phenomenon.

\subsection{Probe physics and spacetime singularity resolution}

The physics seen by a probe in string theory depends on the probe and
the target.  If we probe a target made of fundamental strings with
another string, the best spatial resolution possible turns out to be
the string scale $\ls$.  This happens because the string is an
extended object; more energy-momentum pumped into the probe string
does not result in greater position-space resolution but rather in
stretching the probe.  The existence of a minimum distance is related
to the T-duality symmetry of string theory (for a spatial direction
compactified on a circle of radius $R$, T-duality exchanges
$R/\ls\lra\ls/R$).  If a D-brane is used as a probe of other D-branes,
different physics results.  For example, in the case of D0-branes,
Matrix Theory (see e.g. Ref.~\cite{tomtasi99}) gives a characteristic
scale of $\gs^{1/3}\ls$.

In our situation we are interested in probing the system of $N$
D6-branes wrapped on the $K3$.  It turns out that the best probe for
answering our question about singularity resolution is a clone, namely
a single D6-brane also wrapped on the $K3$.  

We begin with the spacetime or bulk side of the story.  We take
large-$N$, so that the probe D6 can be thought of as a `test'-brane,
and the $N$ `source'-branes are represented by their supergravity
solution.  By supersymmetry, the static potential between the source
and probe branes vanishes.  The action for the probe brane turns out
to be
\be
S = \int
\left\{ \half {\vec{v}}^2 {{R^4}\over{(2\pi)^2\gs\ls^7}}
        {{\Bigl[1-2(\lambda_2/U)\Bigr]}} \right\} 
        \ + {\cal{O}}({\vec{v}}^4) \,,
\ee
where ${\vec{v}}$ is the velocity of the brane in the $(U,\Omega_2)$
directions.  The coefficient of $v^iv^j$ is the metric on moduli
space, and we see that for the D-branes wrapped on $K3$ it is not flat
as it would have been for branes wrapped on T$^4$.  The moduli space
metric has a zero, which signals the vanishing of the `local' tension
of the probe.  The locus of this zero is a sphere of radius
\be\label{enhrad}
\Ue \equiv 2\lambda_2 =2\gYM^2 N\,.
\ee
Notice that the radius $\Ue$ is {\em{twice}} as far out as the radius
of the repulson singularity.  We may now ask what physics is signified
by the vanishing of the local tension of the probe.  By inspection of
the metric and dilaton, we find that nothing special happens at $\Ue$.
On the other hand, the volume of $K3$, which varies with $U$ as
\be
{\rm{Vol}}(K3) = \ls^4 {{[1-(\lambda_2/U)]}\over{(\lambda_2/U)}} \,,
\ee
goes to the special value $\ls^4$ at $\Ue$.

It is perhaps easiest to interpret the physics by performing T- and
S-dualities to turn the D6-brane wrapped on a $K3$ into a heterotic
string wrapped on a circle.  The D6- and D2-brane charges (+1,-1) turn
into winding and momentum charges (+1,-1), and the $\ls^4$-sized $K3$
turns into a $\ls$-sized circle.  From perturbative string theory it
is known that such strings wound on such a circle are massless and
provide the gauge bosons for an enhanced $SU(2)$ symmetry.
Alternatively, in the dual realization with D3-branes strung between
two NS5-branes, the $SU(2)$ is that of the two NS5-branes, which is
restored at the radius $\Ue$ because brane bending due to the D3's
causes the NS5's to touch there.  Dualizing back to our original
system, we still have the $SU(2)$ enhanced symmetry, at the locus
$U\!=\!\Ue$, which we dub the `enhan{\c{c}}on'.  This $SU(2)$ symmetry
is broken at any distance $U>\Ue$; this is a Higgs mechanism in
disguise.

We can also use the heterotic dual picture to understand physically
why there is no notion of `inside the enhan{\c{c}}on' for the probe
brane; it is just the minimum-distance phenomenon we mentioned at the
beginning of this section.  We can also compute the Compton wavelength
of the probe as it approaches the enhan{\c{c}}on, and we find that it
expands smoothly upon approach to the enhan{\c{c}}on locus.  We refer
the reader to Ref.~\cite{JPP} for details and further explanation.
 
Now let us imagine trying to build the singular repulson geometry, one
brane at a time.  Notice that the radius of the enhan{\c{c}}on locus,
$\Ue$ of Eq.~(\ref{enhrad}), is linear in $N$.  Thus, the first brane
has a small enhan{\c{c}}on radius.  The second brane cannot go inside
the enhan{\c{c}}on radius of first brane, and so we get a pair of
branes at finite separation.  The third brane cannot go inside the
enhan{\c{c}}on radius of the first two, and so on; in this way a
sphere of $N$ evenly spaced branes is built up.  The source branes
then form a `Dyson sphere' of a radius twice that at which the
classical naked singularity occurred.  Therefore, the singularity is
excised - in quantum string theory it was never really there.  The
gravitational field is flat inside the Dyson sphere, by symmetry.
Notice also that at large 't Hooft coupling this Dyson sphere is
macroscopically large.

\subsection{The enhan{\c{c}}on phenomenon and Seiberg-Witten theory}

In the last subsection we saw that stringy effects saved the classical
spacetime from embarrassment by excising its naked singularity.  We
now proceed to exhibit the corresponding phenomenon on the gauge
theory side of the story.

For the analogue of our singularity excision in classical gravity, we
must look to nonperturbative gauge theory.  Fortunately, all we need
for study of the moduli space physics is the Seiberg-Witten (S-W)
curve.  For the case $N\!=\!2$, in moduli space there are two branch
points possessing enhanced gauge symmetry, namely where the monopole
and dyon of the theory become massless.  The large-$N$ version of the
SW curve\footnote{We have switched to the $p\!=\!3$ case for
convenience.} was worked out in Ref.~\cite{SWn},
\be
y^2= \prod_{i=1}^N (x-\phi_i)^2 - \Lambda^{2N} \,,
\ee
where $\Lambda$ is a nonperturbatively generated scale and the
$\phi_i$ are the adjoint scalar field vevs.

\begin{figure}
\epsfysize=0.75in
\hskip0.3\textwidth\epsfbox{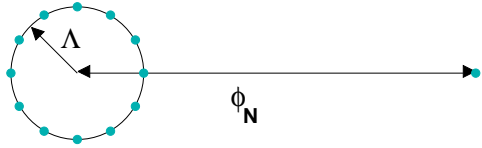}
\caption{\small The brane positions as obtained by studying the
Seiberg-Witten curve.}
\label{fig:seiwit}
\end{figure}

In the situation of interest, namely the system of a single probe
brane far away from $(N\!-\!1)$ other branes, the vevs are
\be
\phi_{i} = 0 \,,\  i=1\ldots N\!-\!1\,, \quad
\phi_N \gg \Lambda \,.
\ee
Solving for the branch points $y=0$ gives $2(N\!-\!1)$ points $x$ on
an $S^1$ of radius $\Lambda$, and two at $x=\phi_N$ (to
${\cal{O}}(1/N)$).  More generally, little is known about
Seiberg-Witten theory for the $d\!=\!p\!+\!1$ gauge theories, but by
analogy we will obtain a $S^{4-p}$ of radius $\Lambda$.  This gives
the brane positions as shown in Fig.\ref{fig:seiwit}.

We can also use the S-W curve to deduce the physics if we try to bring
the probe inside the sphere.  For this, we set
$\phi_N < \Lambda$.
By solving for the branch points in this case, we find that they all
lie on the sphere.  The conclusion we can draw about the physics is
that as we try to adiabatically move the probe brane `inside' the
sphere, it actually smoothly melts into the sphere.

It is worth noting at this point that there are other solutions for
the branch points, but we used a symmetry argument by analogy with the
bulk computation to pin down the spherically symmetric one.  This
suggests that there are other configurations on the bulk side which
correspond to the other solutions from the SW curve, or that there is
a degeneracy to explore.

Although we do not have space to discuss the details here, we have
constructed the phase diagram of the $d\!=\!p\!+\!1$ systems with
eight supercharges.  This turns out to be significantly less
straightforward than for sixteen supercharges.  One reason is that
there is a region of the phase diagram which does not have an obvious
weakly coupled dual.  This is the region describing energies of the
order of the masses of strings stretched between different branes on
the enhan{\c{c}}on sphere. 
In addition, the finite temperature version of our setup does not
permit black hole horizons.  For the details, and some suggestions
about the nature of the missing component of the phase diagram, we
refer the reader to Ref.~\cite{JPP}.  

Ref.~\cite{clifford} shows that the enhan{\c{c}}on phenomenon also
appears in $SO(2N\!+\!1)$, $USp(2N)$ and $SO(2N)$ gauge theories.

In the future we expect to develop further the physics of the
enhan{\c{c}}on phenomenon, and to make links with other recent studies
of singularity resolution in string theory, such as
Ref.~\cite{goodbadnaked} and Ref.~\cite{joematt}.

\bigskip\bigskip\bigskip
\section*{Acknowledgments}

The author wishes to acknowledge co-authors on [1], and in addition
helpful discussions with Eric D'Hoker and Gary Horowitz.  

\medskip\noindent
This work was supported in part by NSF grant PHY94-07194.


\end{document}